\documentclass[final,twoside,journal]{IEEEtran}
\usepackage{subfigure}
\usepackage{graphicx}
\usepackage{amssymb}

\newcommand{\boldsymbol}[1]{\mbox{\boldmath $#1$}}
\def\vec#1{{\boldsymbol #1}}

\begin{document}

\title{Boundary-integral method for poloidal axisymmetric AC magnetic fields}
\author{J\={a}nis~Priede~and~Gunter~Gerbeth%
\thanks{Manuscript received April 28, 2004; revised October 27, 2005.
This work was supported by Deutsche Forschungsgemeinschaft in frame
of the Collaborative Research Centre SFB 609 and by the European Commission
under grant No. G1MA-CT-2002-04046.}%
\thanks{J. Priede was with Forschungszentrum Rossendorf, P.O. Box 510119,
01314 Dresden, Germany during this work. Presently he is with the
Institute of Physics, University of Latvia, Miera st. 32, LV--2169
Salaspils, Latvia; \protect \\
\indent G. Gerbeth is with Forschungszentrum Rossendorf, MHD Department,
P.O. Box 510119, 01314 Dresden, Germany}}

\markboth{IEEE Transactions on Magnetics,~Vol.~42, No.~2,~February~2006}%
{Priede and Gerbeth: Boundary-integral method for poloidal axisymmetric AC magnetic fields}
\pubid{0018--9464/\$20.00~\copyright~2005 IEEE}

\maketitle
\begin{abstract}
This paper presents a boundary-integral equation (BIE) method for
the calculation of poloidal axisymmetric magnetic fields applicable
in a wide range of ac frequencies. The method is based on the vector
potential formulation and it uses the Green's functions of Laplace
and Helmholtz equations for the exterior and interior of conductors,
respectively. The work is particularly focused on a calculation of
axisymmetric Green's function for the Helmholtz equation which is
both simpler and more accurate compared to previous approaches. Three
different approaches are used for calculation of the Green's function
depending on the parameter range. For low and high dimensionless ac
frequencies we use a power series expansion in terms of elliptical
integrals and an asymptotic series in terms of modified Bessel functions
of second kind, respectively. For the intermediate frequency range,
Gauss-Chebyshev-Lobatto quadratures are used. The method is verified
by comparing with the analytical solution for a sphere in a uniform
external ac field. The application of the method is demonstrated for
a composite model inductor containing an external secondary circuit.
\end{abstract}


\section{Introduction}

\PARstart{C}{alculation} of alternating magnetic fields and the associated 
eddy currents is important for the design of various electrical machines
and the magnetic field inductors used for heating, melting, stirring,
shaping or levitation of metallic or semiconducting materials. Although
the distribution of electromagnetic fields is, in principle, completely
described by the Maxwell equations, only in very few simple cases
these equations can be solved analytically. Usually a numerical approach
is needed.

Most of the approaches used for the solution of electromagnetic
field problems are based on finite difference (FDM) or finite
element methods (FEM). The main advantage of these methods is
their capability to deal with complex geometrical configurations
usually encountered in practical applications. However, these
methods involve a solution for the fields in the free space which
is often unbounded and, thus, may require considerable additional
computer resources as well as a special numerical treatment at the
outer open boundary \cite{Gratkowskietal-2000}. On the other hand,
in many applications such as, for example, electromagnetic heating
and hardening of workpieces or the stirring of molten metals, only
the eddy currents and magnetic fields in the conducting medium are
needed. There are several approaches avoiding the solution of the
magnetic field in the free space. A first kind of those approaches
uses the Biot-Savart law, to reduce the problem to a volume
integral equation
\cite{Chitarinetal-1989,KettunenForsman-1996,VolzMazuruk-2003}. A
second type of approaches combine FEM for the solution of the
corresponding partial differential equation (PDE) inside the
conductor with a boundary element method (BEM) based on the second
Green's theorem to represent the field in the free space as an
integral of the field values and its gradient over the surface of
the conductor \cite{SongLi-1999,BoermOstrowski-2003}. In case of a
thin skin-layer this approach reduces to a boundary integral
equation \cite{Tsuboietal-1994}. On the other hand, there are
approaches using the boundary impedance condition to approximate
the field distribution in the conductor at small skin-depth in
combination with a FEM solution in the free space
\cite{DumontGagnoud-2000}.

\pubidadjcol

Our approach is to reduce the problem to a boundary integral
equation not only for the free space but also for the interior of
the conductor that would be applicable regardless of the relative
thickness of the skin-layer. This is possible for a conductor with
uniform electric and magnetic properties when the field
distribution inside the conductor is described by linear PDEs with
constant coefficients admitting an analytic fundamental solution,
\textit{i.e.,} the Green's function. Advantage of this approach is
a consideration of the conductor surface only. Thus, the
dimensionality of the problem is reduced by one that renders this
method particularly suited for the analysis of complicated
geometrical configurations. On the other hand, this geometrical
simplification comes at the price of an increased algebraic and
numeric complexity due to the calculation of the axisymmetric
Green's function. Similar approaches have already been considered
for 2D \cite{Schneider-Salon-1980,Basso-Ndjock-Broche-1988} and 3D
\cite{KimAliWhite-1993} problems which both are considerably
algebraically simpler than the axisymmetric problem considered
here. There is an analytic solution for the Green's function for
the 2D case and a point-source Green's function is used in the 3D
case while there is no simple analytic solution for Green's
function in the axisymmetric case. The axisymmetric case has been
addressed by Fawzi \textit{et al.} \cite{Fawzi} who derived
boundary integral equations for the transverse magnetic (TM) mode
in terms of the azimuthal electrical and tangential magnetic
fields in the full electrodynamic formulation including the
displacement current. The same problem has been revisited in Refs.
\cite{Huang,Kost} in a quasi-static approximation. Our approach
differs from the previous ones by a more exact calculation of the
Green's function using a combination of analytic, asymptotic and
numeric methods.

The paper is organized as follows. In Section II, problem formulation
and basic equations are given. The boundary-integral equation derivation
and the calculation of the Green's function for the azimuthal component
of the vector potential is presented in Section III. In Section IV, we describe
the numerical implementation of the method and give several application
examples. Finally, summary and conclusions are given in Section V.

\section{Problem formulation and basic equations}

Consider an axisymmetric body of a characteristic size $R_{0}$ at
rest having a uniform electrical conductivity $\sigma$ placed in
an axisymmetric external ac magnetic field with induction $\vec{B}$
alternating harmonically in time with the angular frequency $\omega.$
Searching for magnetic and electric fields in terms of vector and
scalar potentials as 
$\vec{A}\left(\vec{r},t\right)=\Re\left[\vec{A}(\vec{r})e^{i\omega t}\right]$
and $\Phi\left(\vec{r},t\right)=\Re\left[\Phi(\vec{r})e^{i\omega t}\right],$
where $\vec{A}(\vec{r})$ and $\Phi(\vec{r})$ are generally complex
axisymmetric amplitudes, leads to the governing equation: 
\begin{equation}
i\omega\vec{A}+\frac{1}{\mu_{0}\sigma}\vec{\nabla}\times
\vec{\nabla}\times\vec{A}=-\vec{\nabla}\Phi.
\label{eq:A-amp}
\end{equation}
In Eq. (\ref{eq:A-amp}) the gradient of the scalar potential $\vec{\nabla}\Phi$
plays the role of a source term with respect to the vector potential.
As it will be shown later, this source term can be used to specify
an externally applied ac voltage to an axisymmetric coil system. For
the following it is advantageous to use the transformation 
\begin{equation}
\vec{A}=\vec{A}'+i\omega^{-1}\vec{\nabla}\Phi
\label{eq:A-phi}
\end{equation}
that allows us to remove the source term from Eq. (\ref{eq:A-amp})
by including it into the vector potential. Then the equation for $\vec{A}'$
satisfying the Coulomb gauge $\vec{\nabla}\cdot\vec{A}'=0$ can be written as 
\begin{equation}
\vec{\nabla}^{2}\vec{A}'-\lambda^{2}\vec{A}'=0,
\label{eq:A-unf}
\end{equation}
where 
$\lambda^{2}=i\bar{\omega}$ with $\bar{\omega}=\mu_{0}\sigma\omega R_{0}^{2}$
being the dimensionless frequency. Henceforth all quantities and differential
operators are supposed to be nondimensionalized by using the corresponding
characteristic length and vector potential scales, $R_{0}$ and $A_{0},$
where the latter will be specified in the following for each particular
problem. The advantage of Eq. (\ref{eq:A-unf}) compared to its nonuniform
counterpart (Eq. \ref{eq:A-amp}) is that the solution of the former
can straightforwardly be written as a surface integral which is the
aim of the next section.

\section{Boundary-integral equation for the vector potential}

Using second Green's vector theorem the solution of Eq. (\ref{eq:A-unf})
satisfying the Coulomb gauge can be written as: 
\setlength{\arraycolsep}{0.0em}
\begin{eqnarray}
\vec{A}'(\vec{r})=\frac{1}{4\pi}\int_{S}&&\left[\vec{\nabla}G^{\lambda}
(\vec{r}'-\vec{r})\times\vec{n}'\times\vec{A}'(\vec{r}')\right.\nonumber \\
 &&{}-\vec{\nabla}G^{\lambda}(\vec{r}'-\vec{r})
 \left(\vec{n}'\cdot\vec{A}'(\vec{r}')\right)\nonumber \\
&&\left.{}-G^{\lambda}(\vec{r}'-\vec{r})\vec{n}'\times
 \vec{\nabla}\times\vec{A}'(\vec{r}')\right]d^{2}\vec{r}'
\label{eq:A'-vol}
\end{eqnarray}
\setlength{\arraycolsep}{5pt}%
where $G^{\lambda}(\vec{r})=
\frac{\exp\left(-\lambda\left|\vec{r}\right|\right)}{\left|\vec{r}\right|}$
is the Green's function of the scalar Helmholtz equation. Note that
our approach here differs from that of Huang \textit{et al.} \cite{Huang}
who uses a scalar counterpart of the second Green identity which is
not correct in general but leads to the right result in the special
case of an axisymmetric and purely azimuthal vector potential which
satisfies the Coulomb gauge straightforwardly. To find $\vec{A}'$
at the point $\vec{r}$ inside the volume enclosed by surface $S$
we need the values of both $\vec{A}'$ and 
$\vec{n}'\times\vec{\nabla}\times\vec{A}'$
on $S.$ Usually both these values are unknown and two vector equations
are needed to find them. The first of the equations is obtained by
approaching the observation point $\vec{r}$ to the boundary $S$
that results in: 
\setlength{\arraycolsep}{0.0em}
\begin{eqnarray}
\int_{S}&&\left[\vec{\nabla}G^{\lambda}(\vec{r}'
-\vec{r})\times\vec{n}'\times\vec{A}'(\vec{r}')
-G^{\lambda}(\vec{r}'-\vec{r}) \right.\nonumber\\
&&\left.\vec{n}'\times\vec{\nabla}\times
\vec{A}'(\vec{r}')\right]d^{2}\vec{r}'-2\pi c(\vec{r})\vec{A}'(\vec{r})=0,
\label{eq:A'-srf}
\end{eqnarray}
\setlength{\arraycolsep}{5pt}%
where $c(\vec{r})$ is a geometrical parameter which is equal to
unity for a smooth surface \cite{Brebbiaetal-1984}. The second equation
is obtained by considering the nonconducting space outside the body
where the distribution of the vector potential is governed by a Laplace
equation. The corresponding equation takes the form 
\setlength{\arraycolsep}{0.0em}
\begin{eqnarray}
\int_{S}\left[\vec{\nabla}G^{0}(\vec{r}'-\vec{r})\times\vec{n}\times\vec{A}
-G^{0}(\vec{r}'-\vec{r})\vec{n}\times\vec{\nabla}\times\vec{A}\right]&&d^{2}\vec{r}'\nonumber\\
{}+2\pi c(\vec{r})\vec{A}(\vec{r})=0,&&
\label{eq:A-srf}
\end{eqnarray}
\setlength{\arraycolsep}{5pt}%
where the sign difference at the second term is because of $\vec{n}$
is directed inwards with respect to the region outside the body. Equation
(\ref{eq:A'-srf}) can now be represented back in terms of the original
vector potential $\vec{A}$ and the imposed gradient of the scalar
potential by inverting the transformation given by Eq. (\ref{eq:A-phi}):
$\vec{A}'=\vec{A}-i\omega^{-1}\vec{\nabla}\Phi.$

In the following we focus on the case of a purely azimuthal and axisymmetric
vector potential $\vec{A}(\vec{r})=\vec{e}_{\varphi}A(r,z)$ depending
only on the radius $r$ and the axial coordinate $z$ in a cylindrical
system of coordinates. For the gradient of the scalar potential 
$\vec{\nabla}\Phi$
to be purely azimuthal and axisymmetric, $\Phi$ can be a function
of the azimuthal angle $\varphi$ only: $\Phi=\Phi(\varphi).$ Then
$\vec{\nabla}\Phi=\vec{e}_{\varphi}\frac{1}{r}
\frac{\partial\Phi}{\partial\varphi}$
with $\frac{\partial\Phi}{\partial\varphi}=\Phi_{0}=const$ because
of the axisymmetry. Further note that for axisymmetric bodies with
simply connected shapes including the symmetry axis, $\Phi_{0}$ must
be zero for $\vec{\nabla}\Phi$ to be limited at the symmetry axis
$r=0$. However, $\Phi_{0}$ may be nonzero for toroidal bodies which
are not intersected by the symmetry axis. For such bodies, like coils,
$\Phi_{0}$ may be used to specify the externally applied voltage
$U$ driving the current as 
$\frac{\partial\Phi}{\partial\varphi}=\frac{U}{2\pi}.$
Alternatively, $\Phi_{0}$ may be determined in the course of solution
when the total current rather than the voltage is specified on the
coil. Note that our treatment of the source term is more mathematically
rigorous compared to Ref. \cite{Huang}.

Substituting such a purely azimuthal vector potential into Eq. (\ref{eq:A-srf})
and performing the integration along the azimuthal angle $\varphi$
we obtain after some transformations an equation defining $A$ outside
the conducting body 
\setlength{\arraycolsep}{0.0em}
\begin{eqnarray}
A(\vec{r})=-\frac{1}{4\pi}\int_{L}&&\left[\frac{\partial\left(r'A(\vec{r}')
\right)}{\partial n'}G_{\varphi}^{0}(\vec{r},\vec{r}')\right.\nonumber\\
&&\left.{}-A(\vec{r}')\frac{\partial\left(r'G_{\varphi}^{0}(\vec{r},\vec{r}')
\right)}{\partial n'}\right]d\left|\vec{r}'\right|,
\label{eq:biq-out}
\end{eqnarray}
\setlength{\arraycolsep}{5pt}%
where the integral is now evaluated along the contour $L$ forming
the conducting body of rotation. The Green's function for the azimuthal
component of the vector potential entering the above equation is 
\setlength{\arraycolsep}{0.0em}
\begin{eqnarray}
G_{\varphi}^{0}(\vec{r},\vec{r}') &{}={}& \vec{e}_{\varphi}\cdot
\int_{0}^{2\pi}\vec{e}'_{\varphi}G^{0}(\vec{r}'-\vec{r})d\varphi'\nonumber \\
&{}={}&\frac{2k}{\sqrt{r'r}}\int_{0}^{\pi/2}\frac{2\sin^{2}\varphi-1}
{\sqrt{1-k^{2}\sin^{2}\varphi}}d\varphi\nonumber \\
&{}={}& \frac{4k}{\sqrt{r'r}}\left[\frac{K(k)-E(k)}{k^{2}}
-\frac{K(k)}{2}\right],
\label{eq:g-0}
\end{eqnarray}
\setlength{\arraycolsep}{5pt}%
which is the vector potential of a circular current loop divided
by $r'$ \cite{Jackson-1975} presented in terms of the complete elliptical
integrals of the first and second kind, $K(k)$ and $E(k)$, respectively,
of the modulus $k=2\sqrt{\frac{r'r}{(r'+r)^{2}+(z'-z)^{2}}}$ \cite{Abramowitz}.
Thus, the Green's function like its gradient for the azimuthal component
of the Laplace equation is obtained analytically.

The azimuthal component of the vector potential inside the conducting
body can be obtained in a similar way as outside by using the corresponding
Green's function with $\lambda\neq0$
\setlength{\arraycolsep}{0.0em}
\begin{eqnarray}
A'(\vec{r})=\frac{1}{4\pi}\int_{L}&&\left[\frac{\partial\left(r'A'(\vec{r}')
\right)}{\partial n'}G_{\varphi}^{\lambda}(\vec{r},\vec{r}')\right.\nonumber\\
&&\left.{}-A'(\vec{r}')\frac{\partial\left(r'G_{\varphi}^{\lambda}(\vec{r},\vec{r}')
\right)}{\partial n'}\right]d\left|\vec{r}'\right|,
\label{eq:biq-in}
\end{eqnarray}
\setlength{\arraycolsep}{5pt}%
where 
\setlength{\arraycolsep}{0.0em}
\begin{eqnarray}
G_{\varphi}^{\lambda}(\vec{r},\vec{r}') &{}={}& \vec{e}_{\varphi}\cdot
\int_{0}^{2\pi}\vec{e}'_{\varphi}G^{\lambda}(\vec{r}'-\vec{r})d\varphi'\nonumber \\
&{}={}& \frac{2k}{\sqrt{r'r}}\int_{0}^{\pi/2}\frac{2\sin^{2}\varphi-1}
{\sqrt{1-k^{2}\sin^{2}\varphi}}\nonumber\\
&&\qquad\times\exp\left(-\kappa\sqrt{1-k^{2}\sin^{2}\varphi}\right)d\varphi,
\label{eq:g-lambda}
\end{eqnarray}
\setlength{\arraycolsep}{5pt}%
and $\kappa=2\lambda\sqrt{r'r}/k.$ In contrast to the previous case
with $\lambda=0$, the last integral cannot be evaluated analytically.
For $\left|\kappa\right|\ll1,$ corresponding to low frequencies,
the exponential function in (\ref{eq:g-lambda}) may be expanded into
a power series of $\kappa:$
\begin{equation}
G_{\varphi}^{\lambda}(\vec{r},\vec{r}')=-\frac{2k}{\sqrt{r'r}}\sum_{n=0}^{\infty}
\frac{(-\kappa)^{n}}{n!}\left(I_{n}+\frac{4}{n+1}I_{n}'\right),
\label{eq:g-pwr}
\end{equation}
where $I_{n}=\int_{0}^{\pi/2}\left(1-k^{2}\sin^{2}\varphi\right)^{\frac{n-1}{2}}
d\varphi=\left\{ \begin{array}{ll}
I_{l}^{o}, & n=2l+1\\
I_{l}^{e}, & n=2l\end{array}\right.,\quad l=0,1,2,..$, and 
$I_{n}'=\frac{dI_{n+1}}{dk^{2}}$. For odd $n$ the theory of
elliptical integrals \cite{Korn} yields the following recursion
\[
I_{l+2}^{o}=\frac{2l+3}{2l+4}(2-k^{2})I_{l+1}^{o}
-\frac{l+1}{l+2}(1-k^{2})I_{l}^{o},
\]
with $I_{0}^{o}=\frac{\pi}{2}$ and $I_{1}^{o}=\frac{\pi}{4}(2-k^{2})$.
Derivative of this recursion with respect to $k^{2}$ leads to a similar
recursion for $I_{n}^{o'}.$ Similarly, for even indices one obtains:
\[
I_{l+2}^{e}=\frac{2l+2}{2l+3}(2-k^{2})I_{l+1}^{e}
-\frac{2l+1}{2l+3}(1-k^{2})I_{l}^{e},
\]
with $I_{0}^{e}=K(k)$ and $I_{1}^{e}=E(k)$. Series (\ref{eq:g-pwr})
is summed until $\frac{\left|\kappa\right|^{n}}{n!}<10^{-8}$ that
ensures a relative error less than $10^{-5}$ for $\left|\kappa\right|<5k^{2}.$

At high frequencies, when $\left|\kappa\right|\gg1,$ (\ref{eq:g-lambda})
is dominated by the maximum of the exponential function about the
point $\varphi=\frac{\pi}{2}$ and it is possible to evaluate it asymptotically
by the Laplace method \cite{Hinch}. Substitution of $\cos\varphi=t$
in Eq. (\ref{eq:g-lambda}) results in 
\begin{equation}
G_{\varphi}^{\lambda}(\vec{r},\vec{r}')=\frac{2\beta}{\sqrt{r'r}}
\int_{0}^{1}\frac{\exp\left(-s\sqrt{1+\beta^{2}t^{2}}\right)}
{\sqrt{1+\beta^{2}t^{2}}}\frac{\left(1-2t^{2}\right)}{\sqrt{1-t^{2}}}dt,
\label{eq:g-sub}
\end{equation}
where $s=\kappa\sqrt{1-k^{2}}$ and $\beta=\frac{k}{\sqrt{1-k^{2}}}.$
Since the dominating contribution in the above integral results from
the vicinity of $t=0$ we can expand 
$\frac{1}{\sqrt{1-t^{2}}}=
\sum_{m=0}^{\infty}\frac{\Gamma(m+1/2)}{\sqrt{\pi}m!}t^{2m}$
and shift the upper limit of integration to infinity 
\setlength{\arraycolsep}{0.0em}
\begin{eqnarray}
\nonumber
G_{\varphi}^{\lambda}(\vec{r},\vec{r}') &{}={}& \frac{2}{\sqrt{r'r}}
\int_{0}^{\infty}\exp\left(-s\cosh x\right)\left(1-2\left(\frac{\sinh x}
{\beta}\right)^{2}\right)\nonumber\\
&&\qquad{}\times\sum_{m=0}^{\infty}\frac{\Gamma(m+1/2)}{\sqrt{\pi}m!}
\left(\frac{\sinh x}{\beta}\right)^{2m}dx\nonumber \\
&{}={}& \frac{2}{\sqrt{r'r}}\sum_{m=0}^{\infty}\frac{\Gamma(m+1/2)}
{\sqrt{\pi}m!\beta^{2m}}\left(I_{m}-\frac{2}{\beta^{2}}I_{m+1}\right),
\label{eq:g-asm}
\end{eqnarray}
\setlength{\arraycolsep}{5pt}%
where we have made the additional substitution $t=\frac{\sinh x}{\beta}$.
The integrals in the above relation 
\setlength{\arraycolsep}{0.0em}
\begin{eqnarray*}
I_{m}&{}={}&\int_{0}^{\infty}\exp\left(-s\cosh x\right)\sinh^{2m}xdx\\
&{}={}&\frac{\Gamma(m+1/2)}{\sqrt{\pi}}\left(\frac{2}{s}\right)^{m}K_{m}(s),
\end{eqnarray*}
\setlength{\arraycolsep}{5pt}%
defined in terms of the modified Bessel function of the second kind
of order $m$, $K_{m}(s),$ \cite{Abramowitz}, can efficiently be
calculated for $m>1$ by the following recursion: 
\[
I_{m+1}=(2m+1)\left(2mI_{m}+(2m-1)I_{m-1}\right)/s^{2}.\]
 The gradient of $G_{\varphi}^{\lambda}$ can be found in a similar
way by using the relation $\frac{dI_{m}}{ds}=-\frac{sI_{m+1}}{2m+1}$
which follows from the properties of Bessel functions \cite{Abramowitz}.

There is an additional range of parameters where the power series
solution given by Eq. (\ref{eq:g-pwr}) is not applicable because
$\left|\kappa\right|$ is large, while the asymptotic approximation
(\ref{eq:g-asm}) does not work because $k$ is small and the exponential
function under the integral (\ref{eq:g-lambda}) varies weakly along
the angle $\varphi$ without having a pronounced maximum. In this
case, one could expand the sub-integral function in Eq. (\ref{eq:g-lambda})
in a power series of $k^{2}.$ As easy to see, this would result in
the power series of $\sin^{2}\varphi$ which can in principle be integrated
analytically term by term. On the other hand, such polynomials can
efficiently be integrated by Gauss-Chebyshev-Lobatto quadratures.
Thus, instead of expanding the integral (\ref{eq:g-lambda}) in a
power series of small $k^{2}$ and then integrating analytically term
by term, we apply a Gauss-Chebyshev-Lobatto quadrature \cite{Abramowitz}
directly to the integral (\ref{eq:g-sub}).

To summarize, three different approaches are used for the
evaluation of the Green's function and its gradient for
$\lambda\neq0$ within the following parameter ranges defined in
terms of $k^{2}$ and $\left|\kappa\right|$ which actually specify
the integral in (\ref{eq:g-lambda}). First, for sufficiently small
$\left|\kappa\right|\leq5k^{2}$ we use the power series expansion
(\ref{eq:g-pwr}). Second, for the intermediate range
$5k^{2}<\left|\kappa\right|<35k^{-2}$ a Gauss-Chebyshev-Lobatto
quadrature with $M=30+120k^{4}$ number of points is used where the
number of points is increased as $k\rightarrow1$ in order to
ensure sufficient accuracy in the vicinity of the singularity at
$k=1.$ In addition, for $k^{2}>0.98$ we subtract the singularity
as the zero-frequency Green's function which can be integrated
analytically whereas the rest is integrated numerically as described
above. For $\left|\kappa\right|\geq35k^{-2}$ the first five terms
of the asymptotic series (\ref{eq:g-asm}) are used. The ranges of
applicability of different approximations and the number of the
quadrature points are found numerically and they ensure the
relative error of the Green's function and its gradient to be
below $10^{-5}$ for $k\lesssim0.999.$

Two coupled boundary-integral equations are obtained from Eqs. (\ref{eq:biq-out})
and (\ref{eq:biq-in}) by taking the observation point $\vec{r}$
to the surface contour $L:$
\setlength{\arraycolsep}{0.0em}
\begin{eqnarray}
\int_{L}\left[\frac{\partial\Psi(\vec{r}')}{\partial n'}
rG_{\varphi}^{0}(\vec{r},\vec{r}')-\frac{\Psi(\vec{r}')}{r'}
\frac{\partial\left(r'rG_{\varphi}^{0}(\vec{r},\vec{r}')\right)}
{\partial n'}\right]&&d\left|\vec{r}'\right| \nonumber\\
{}-2\pi c(\vec{r})\Psi(\vec{r}) = 0;&& \label{eq:biqs-out}\\
\int_{L}\left[\frac{\partial\Psi'(\vec{r}')}{\partial n'}
rG_{\varphi}^{\lambda}(\vec{r},\vec{r}')-\frac{\Psi'(\vec{r}')}{r'}
\frac{\partial\left(r'rG_{\varphi}^{\lambda}(\vec{r},\vec{r}')\right)}
{\partial n'}\right]&&d\left|\vec{r}'\right|\nonumber\\
{}+2\pi c(\vec{r})\Psi'(\vec{r})=0,&&\label{eq:biqs-in}
\end{eqnarray}
\setlength{\arraycolsep}{5pt}%
where the unknown functions to be found along $L$ are 
$\Psi(\vec{r})=rA(\vec{r})$
and $\frac{\partial\Psi(\vec{r})}{\partial n},$ while 
$\Psi'(\vec{r})=\Psi(\vec{r})-i\bar{\omega}^{-1}\Phi_{0}$
for the interior involves an additional constant $\Phi_{0}$ defining
the azimuthal gradient of the electrostatic potential which, as discussed
above, may be non-zero for the conductors not intersected by the symmetry
axis. For such conductors the geometrical parameter in Eqs. (\ref{eq:biqs-out},
\ref{eq:biqs-in}) can be determined as 
\begin{equation}
c(\vec{r})=\frac{1}{2\pi}\int_{L}\frac{1}{r'}\frac{\partial\left(
r'rG_{\varphi}^{0}(\vec{r},\vec{r}')\right)}{\partial n'}d\left|\vec{r}'\right|,
\label{eq:biqs-c}
\end{equation}
which follows from the requirement for Eq. (\ref{eq:biqs-in}) to
be satisfied by a constant in the limit of $\lambda\rightarrow0$
similarly to its PDE counterpart (\ref{eq:A-unf}).

\section{Numerical implementation and examples of application}

\begin{figure*}
\centering
\includegraphics[width=\columnwidth]{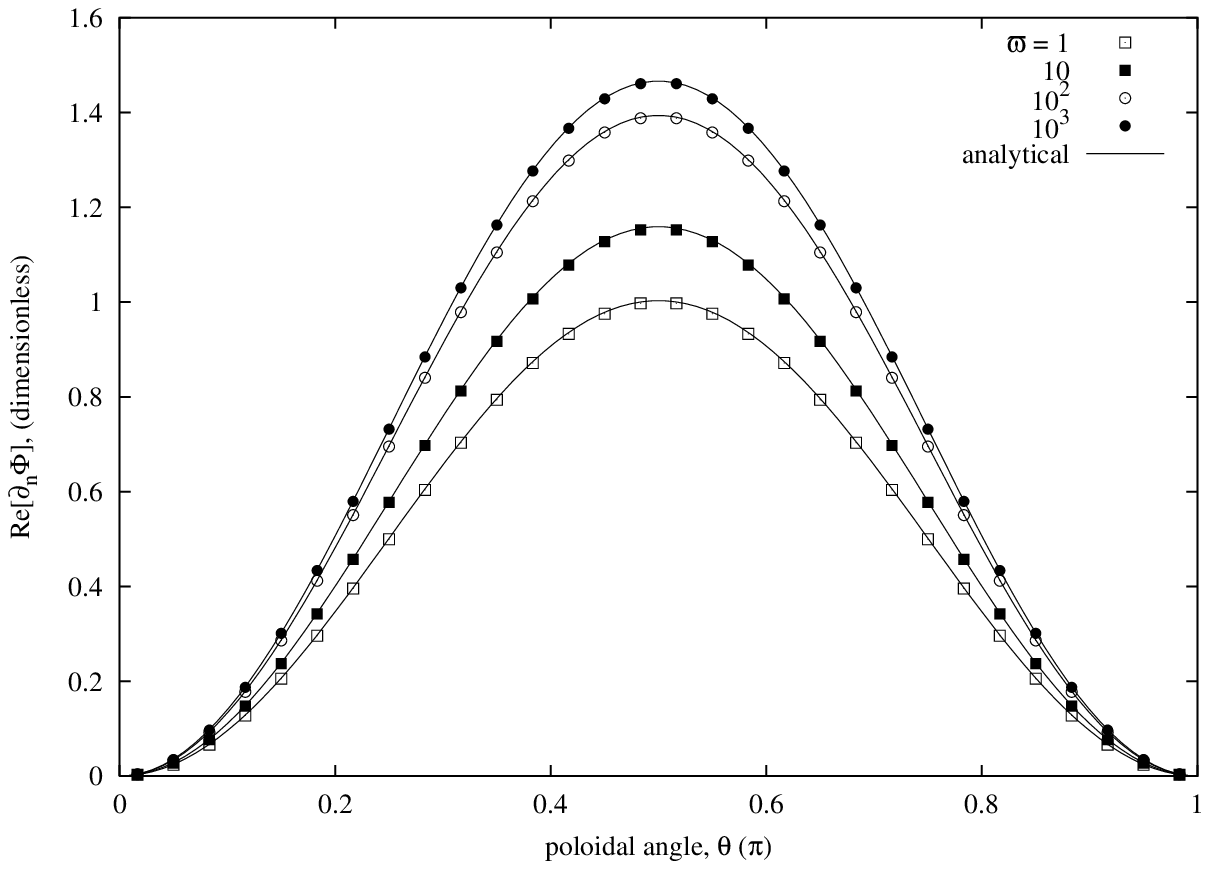}
\includegraphics[width=\columnwidth]{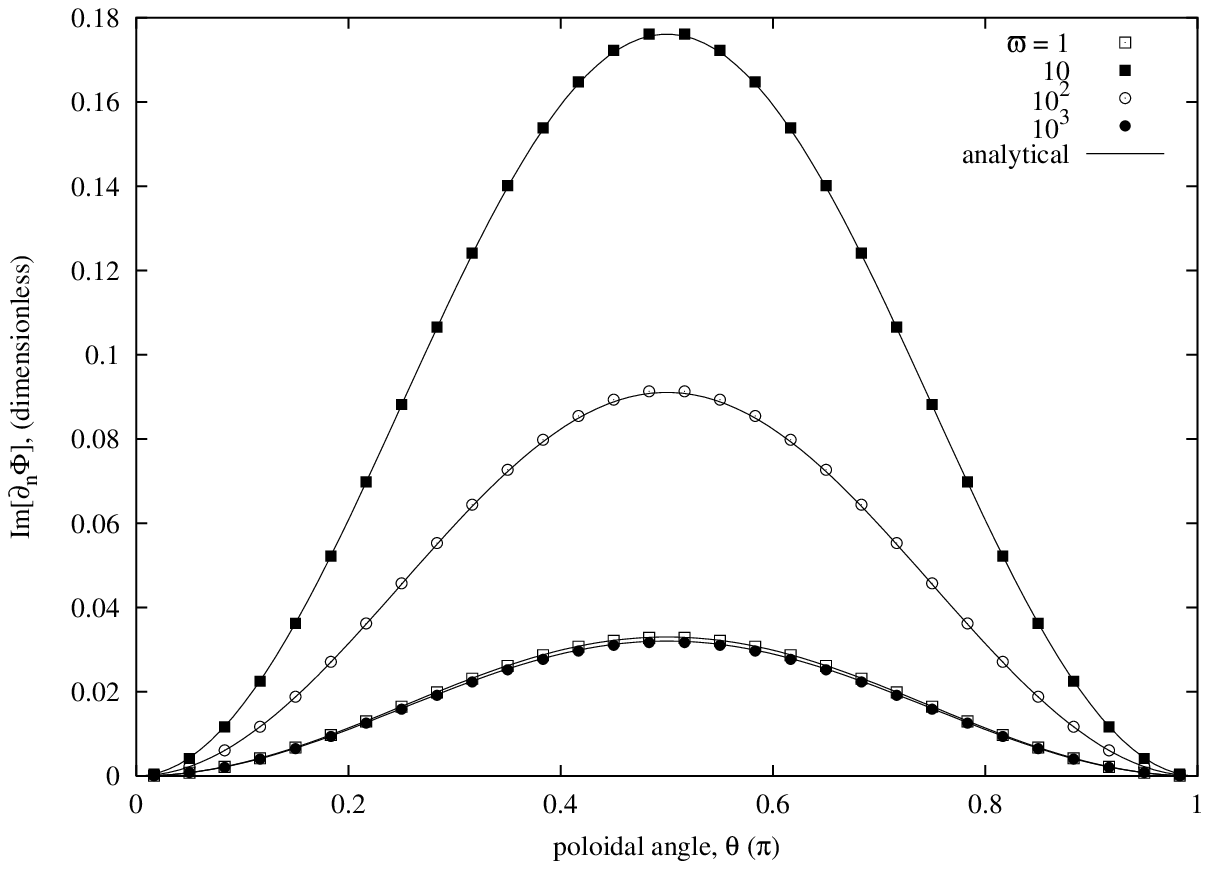}

\centering
\includegraphics[width=\columnwidth]{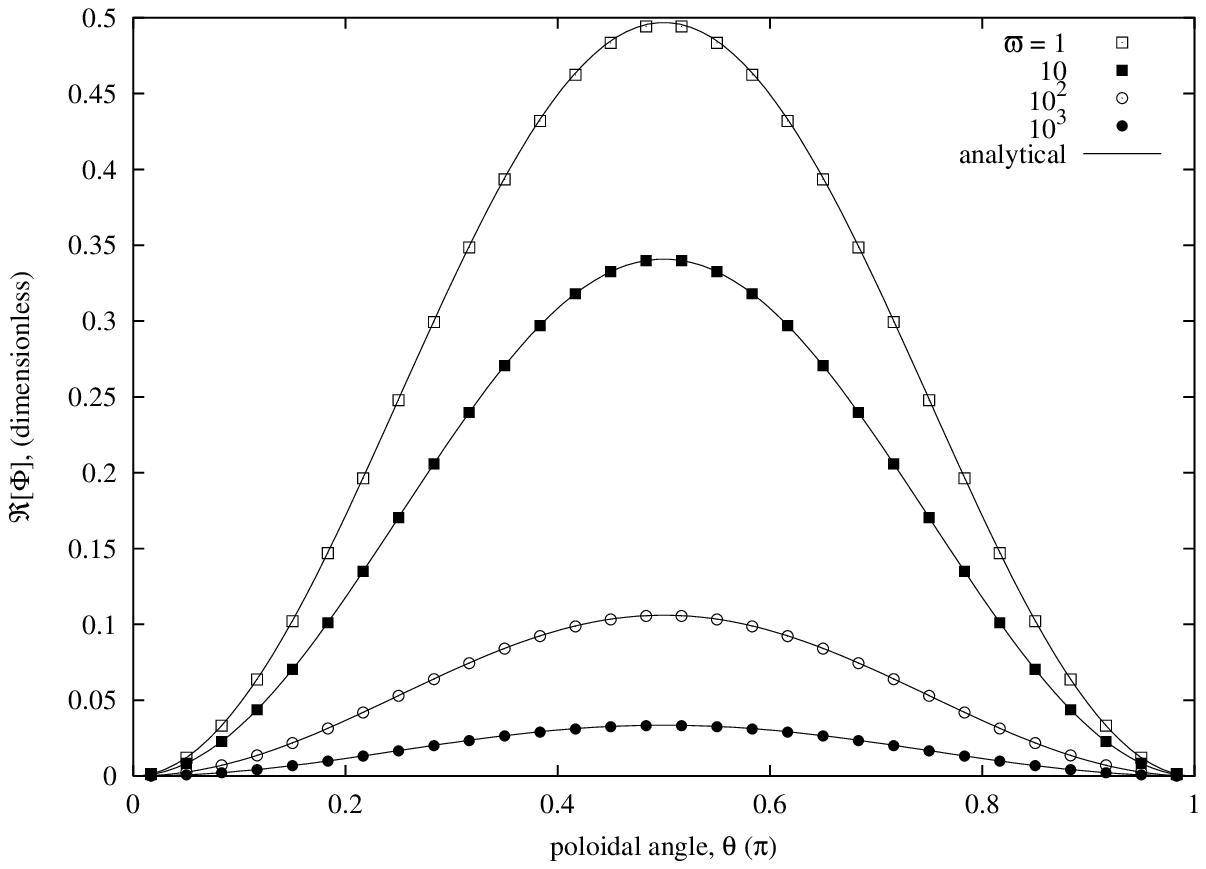}
\includegraphics[width=\columnwidth]{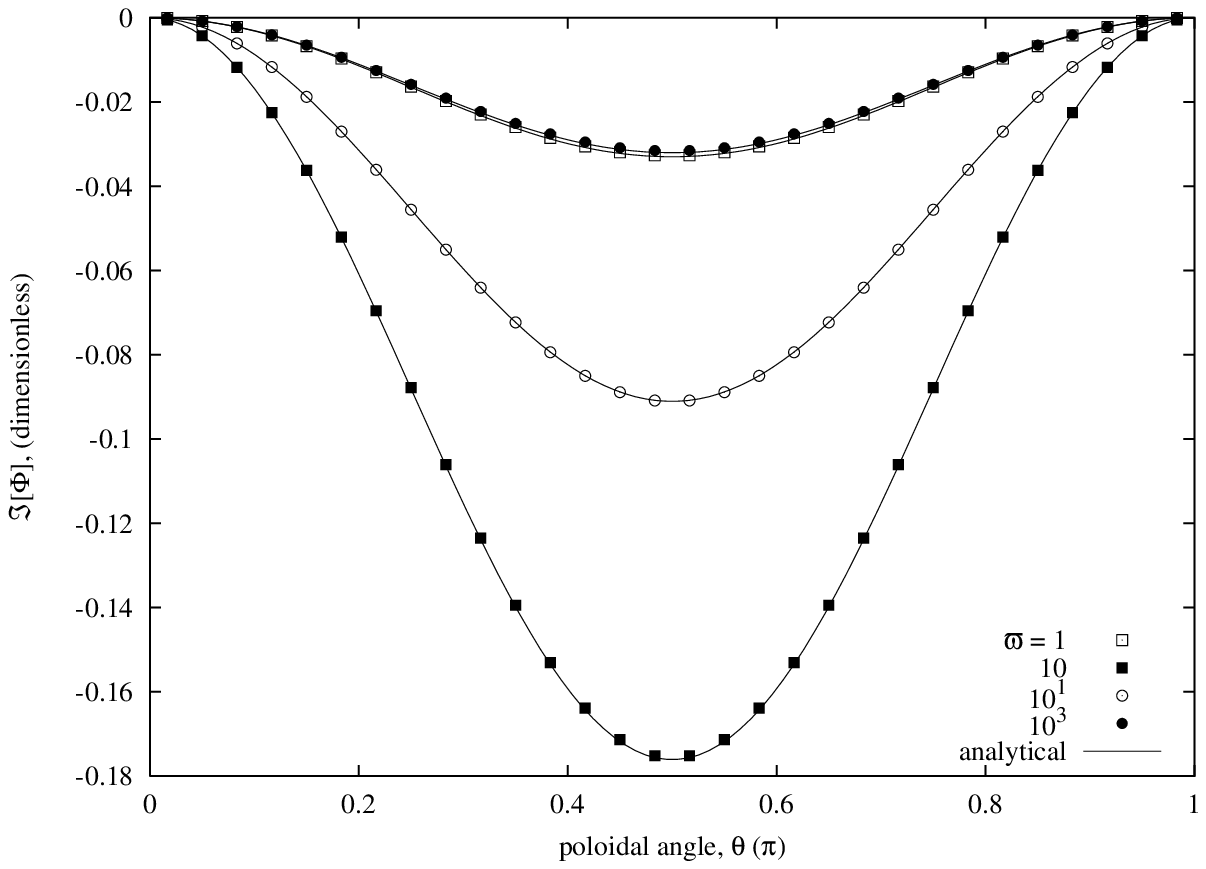}

\caption{\label{cap: sph-sol} Comparison of numerical (dots) and analytical
(solid curves) solutions for a sphere in a uniform ac magnetic field:
real (left) and imaginary (right) parts of $\Psi$ (top) and $\frac{\partial\Psi}{\partial n}$
(bottom) at the surface of the sphere versus the poloidal angle for
various dimensionless ac frequencies. }
\end{figure*}

The system of two coupled boundary-integral Eqs. (\ref{eq:biqs-out},
\ref{eq:biqs-in}) can be solved numerically by the boundary element
method \cite{Brebbiaetal-1984}. For this purpose each line $L_{k}$
forming a closed surface of a part of the conducting body of rotation,
which may be simply or multiply connected, is approximated by $N$
rectilinear segments with endpoints pointed by radius vectors 
$\vec{r}_{i},\, i=1,..,N+1.$
The integrals in Eqs. (\ref{eq:biqs-out}-\ref{eq:biqs-c}) along
each contour are replaced by the sums over the corresponding boundary
elements where the integrals over each boundary element are approximated
by four-point Gauss quadratures \cite{num-recep}. When the observation
and integration points coincide there is a logarithmic singularity
in the Green's function which is subtracted and integrated analytically
over the corresponding element. In the simplest case, the unknown
functions are considered to be constant within each element that results
in $2N$ unknown quantities which are the constant values of $\Psi(\vec{r})$
and $\frac{\partial\Psi(\vec{r})}{\partial n}$ in each element. Upon
evaluation of both Eqs. (\ref{eq:biqs-out}, \ref{eq:biqs-in}) at
the midpoint of each element we obtain a system of $2N$ complex linear
equations. For a typical number of unknowns of about several hundreds
this problem can straightforwardly be solved by an \textit{LU} decomposition.

In the following, we consider two simple examples of application of
the method. The first example is a conducting sphere of radius $R_{0}$
in a uniform external ac magnetic field with induction amplitude $B_{0}$.
In this case, the contour encircling the whole free space in Eq. (\ref{eq:biqs-out})
may be considered to consist of two contours where $L$ encloses the
sphere while the second one encloses some remote inductor creating
a uniform field with $\Psi_{0}(\vec{r})=r^{2}/2$ that corresponds
to the vector potential scaled by $A_{0}=R_{0}B_{0}.$

Thus Eq. (\ref{eq:biqs-out}) for the outer surface of the sphere
takes the form 
\setlength{\arraycolsep}{0.0em}
\begin{eqnarray*}
\int_{L}\left[\frac{\partial\Psi(\vec{r}')}{\partial n'}rG_{\varphi}^{0}
(\vec{r},\vec{r}')-\frac{\Psi(\vec{r}')}{r'}\frac{\partial\left(
r'rG_{\varphi}^{0}(\vec{r},\vec{r}')\right)}{\partial n'}\right]
d\left|\vec{r}'\right|&&\\
{}+2\pi c(\vec{r})\Psi(\vec{r})=-4\pi\Psi_{0}(\vec{r}),&&
\end{eqnarray*}
\setlength{\arraycolsep}{5pt}%
whereas the corresponding Eq. (\ref{eq:biqs-in}) for the inner surface
remains unchanged. The distributions of $\Psi(\vec{r})$ and 
$\frac{\partial\Psi(\vec{r})}{\partial n}$
calculated with $N=30$ constant surface elements are seen in Fig.
\ref{cap: sph-sol} to be in good agreement with the corresponding
analytical solutions \cite{Smythe}: $\left.\Psi\right|_{\left|\vec{r}\right|=1}=\frac{3}{2}\frac{j_{1}\left(x\right)}{xj_{0}\left(x\right)}\sin^{2}(\theta)$
and $\left.\frac{\partial\Psi}{\partial n}\right|_{\left|\vec{r}\right|=1}=\left(1-\frac{1}{2}\frac{j_{2}\left(x\right)}{j_{0}\left(x\right)}\right)\sin^{2}(\theta)$,
where $x=\sqrt{\bar{\omega}/i},$ $\theta$ is the poloidal angle,
and $j_{n}(x)$ is the spherical Bessel function of order $n$ \cite{Abramowitz}.

\begin{figure*}
\centering
\subfigure[]{\includegraphics[width=\columnwidth]{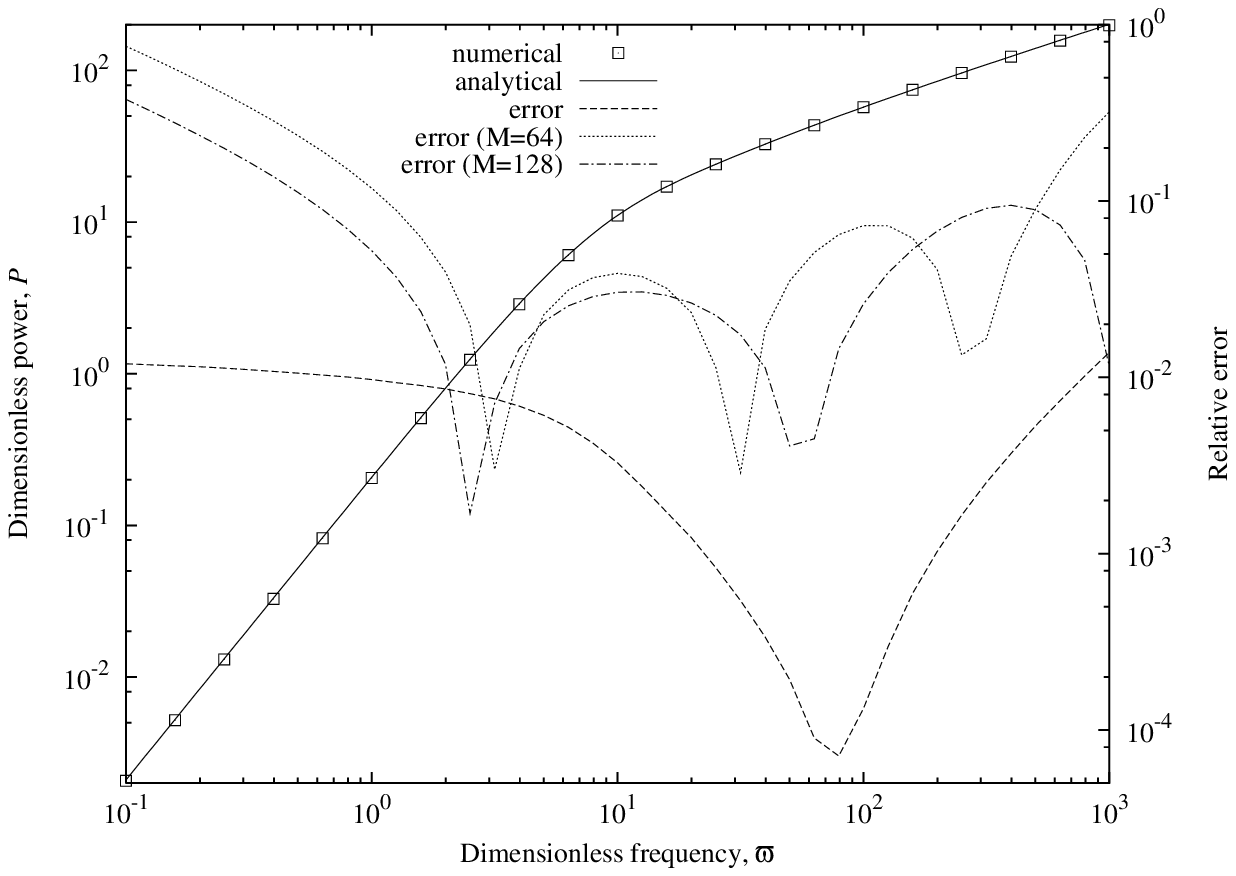}}
\subfigure[]{\includegraphics[width=\columnwidth]{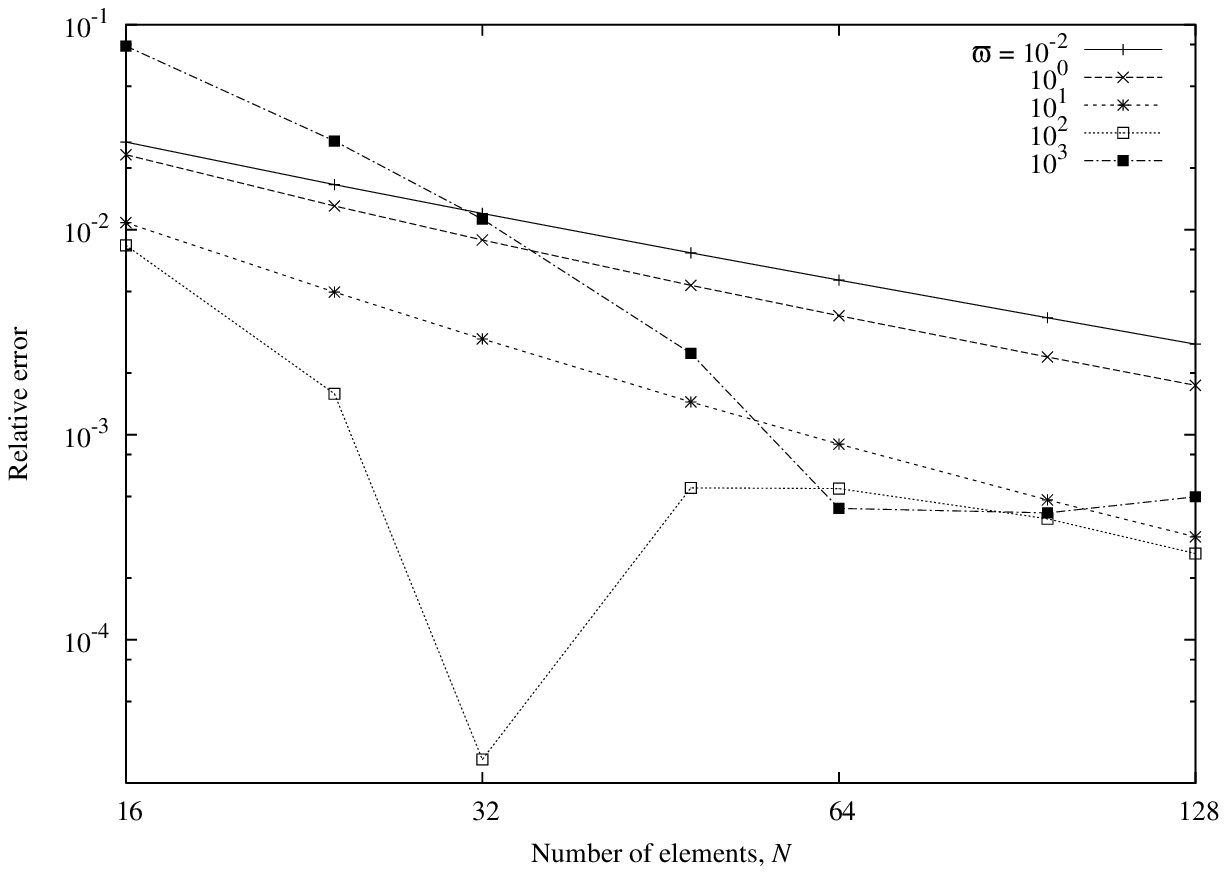}}

\caption{\label{cap:sph-pwr} Comparison of the numerically
obtained dimensionless dissipated power with the exact analytic
solution for a sphere in a uniform ac field: power and the
relative errors resulting from combined analytical/numerical and
purely numerical calculations of the Green's function with $M=64$
and $M=128$ Gauss-Chebyshev-Lobatto quadrature points versus the
dimensionless frequency for $N=30$ boundary elements (a) and
relative error versus the number of BEM at various dimensionless
ac frequencies (b). }
\end{figure*}

An additional quantity which can be used for verification of the method
is the total dissipated power defined in terms of dimensionless surface
quantities as 
\[
P=\pi\bar{\omega}\int_{L}\Im\left[\frac{\partial\Psi}{\partial n}
\frac{\Psi^{*}}{r}\right]d\left|\vec{r}\right|,
\]
where the asterisk denotes the complex conjugate and the power is
scaled by $P_{0}=\frac{R_{o}B_{0}^{2}}{\sigma\mu_{0}^{2}}$. Comparison
of numerical and exact solutions of total power for a sphere in a
uniform ac magnetic field plotted in Fig. \ref{cap:sph-pwr}(a) shows
that 30 constant boundary elements ensure a relative error below a
few per cent for the dimensionless frequency up to $10^{3}.$ For
comparison we show also the relative error of the solution resulting
from purely numerical calculation of the Green's function and its
gradient, as in Ref. \cite{Huang}, with $64$ and $128$ Gauss-Chebyshev-Lobatto
quadrature points that results in a significantly lower accuracy at
both low and high frequencies. As seen in Fig. \ref{cap:sph-pwr}(b),
the accuracy decreases at high frequencies where a larger number of
boundary elements is required. Note that the relatively slow convergence
rate of about $\sim N^{-1}$ is due to the low accuracy of constant
boundary elements used in this example.

\begin{figure*}
\centering
\subfigure[]{\includegraphics[width=\columnwidth]{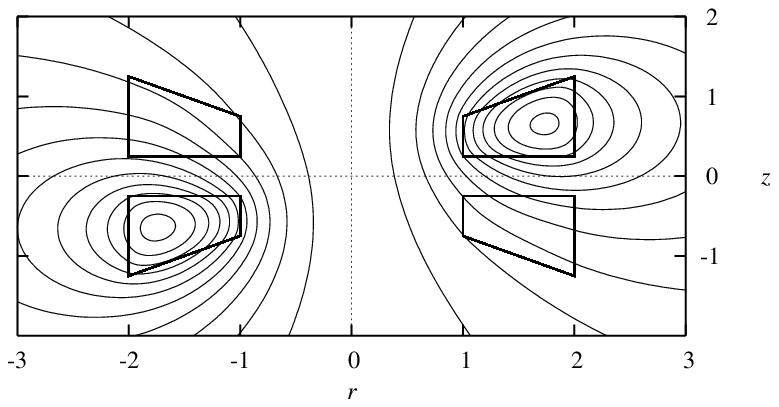}}
\subfigure[]{\includegraphics[width=\columnwidth]{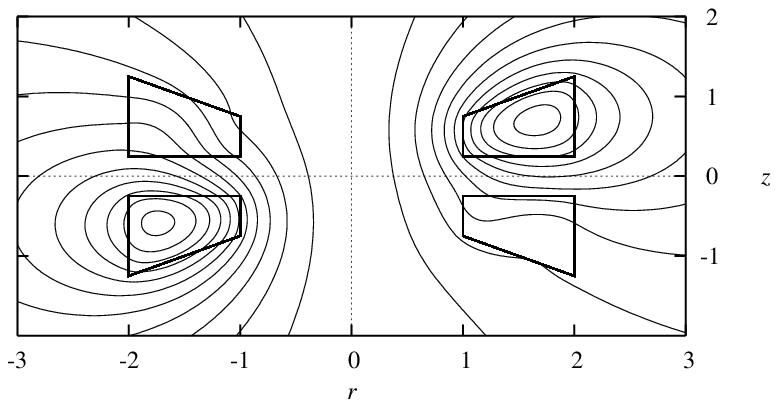}}\\
\subfigure[]{\includegraphics[width=\columnwidth]{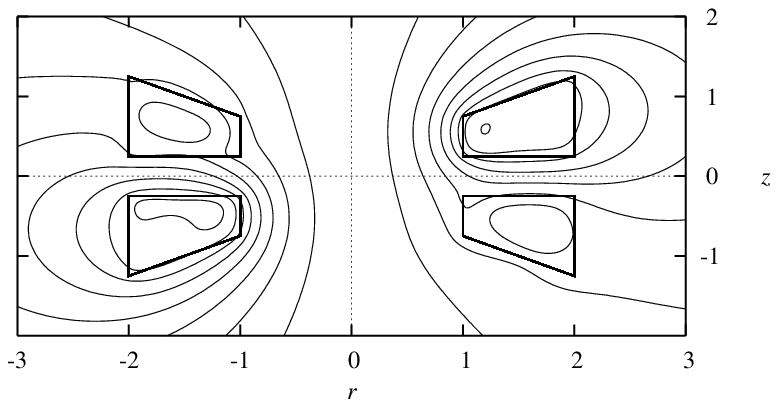}}
\subfigure[]{\includegraphics[width=\columnwidth]{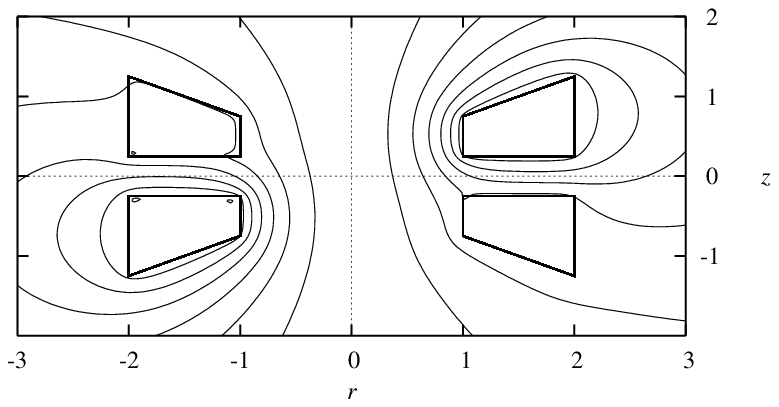}}

\caption{\label{cap:coil-flux} Magnetic flux lines delivered by the surface-integral
equations for a toroidal model inductor at various dimensionless ac
frequencies $\bar{\omega}:$ $1$ (a); $10$ (b); $10^{2}$(c); $10^{3}$(d).
Right and left hand sides of each plot show the magnetic flux lines
in phase and shifted by $\pi/2$ with respect to the current in the
primary (upper) inductor that corresponds to the time instants when
the current is at maximum and zero, respectively. }
\end{figure*}

\begin{figure*}
\centering
\subfigure[]{\includegraphics[width=\columnwidth]{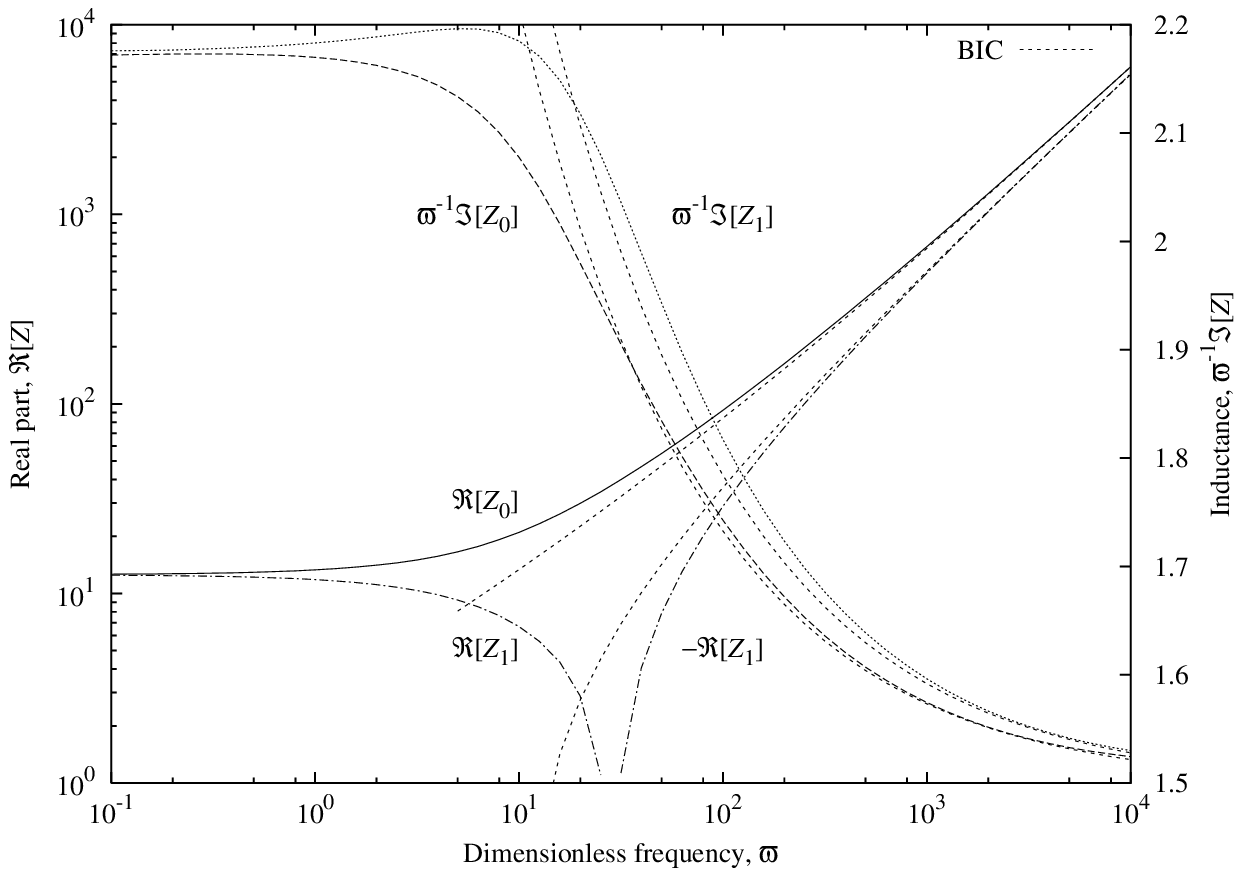}}
\subfigure[]{\includegraphics[width=\columnwidth]{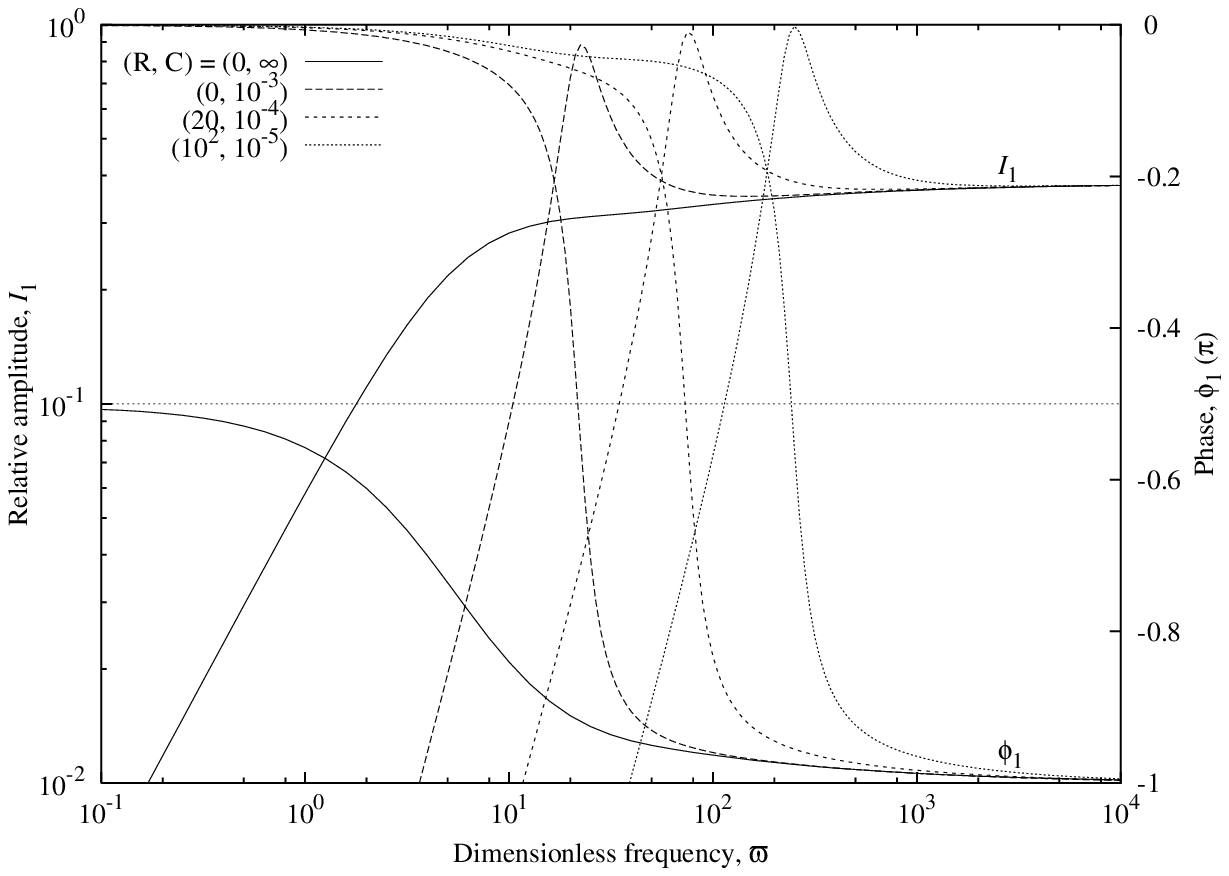}}

\caption{\label{cap:zimped} Dimensionless impedances of primary $(Z_{0})$
and secondary $(Z_{1})$ rings of a model inductor supplied by the
present method and boundary impedance condition (BIC) versus the dimensionless
frequency $\bar{\omega}$ for the secondary current of same amplitude
and $\pi/2$ phase lag with respect to the primary current (\textit{a}).
Relative amplitude and phase of the secondary current at various dimensionless
values of active $(R)$ and capacitive $(C)$ impedance in the secondary
circuit (\textit{b}).}
\end{figure*}

As a next example we consider a model inductor consisting of two coaxial
mirror-symmetric rings of trapezoidal cross-section as shown in Fig.
\ref{cap:coil-flux} related to the crystal growth application by
the floating zone technique \cite{Hermann}. The upper, primary, ring
defined by the contour $L_{0}$ is connected to a power source supplying
ac current $I=I_{0}\cos(\omega t).$ The current in the lower, secondary,
ring $L_{1},$ which is short-circuited through an additional impedance
$Z_{2},$ is induced only by the magnetic field of the upper ring.
In this case, we have two additional unknown quantities, the azimuthal
gradients of the electrostatic potential $\Phi_{0}$ and $\Phi_{1}$
in the primary and secondary inductors. Respectively, we have two
additional equations for the circuits of primary and secondary rings
$I_{0}=\int_{L_{0}}\frac{1}{r}\frac{\partial\Psi(\vec{r})}
{\partial n}d\left|\vec{r}\right|$ and $\int_{L_{1}}\frac{1}{r}
\frac{\partial\Psi(\vec{r})}{\partial n}d\left|\vec{r}\right|
-2\pi\Phi_{1}/Z_{2}=0,$
which are discretized and solved as described above. Our goal here
is to choose the additional impedance $Z_{2}$ so that to have the
current induced in the secondary ring of the same amplitude but delayed
in phase by 90 degrees with respect to the current in the primary
ring. The corresponding magnetic flux lines calculated using 150 equally-sized
boundary elements in each ring are plotted in Fig. \ref{cap:coil-flux}
for various dimensionless frequencies. It has to be noted that the
problem tends to be ill-conditioned at $\bar{\omega}=0$ and, thus,
special care is necessary for calculation of the geometrical parameter
$c(\vec{r})$ defined by Eq. (\ref{eq:biqs-c}) in order to obtain
accurate solutions at low frequencies $\bar{\omega}\ll1$ which, however,
are not very important for practical applications. The dimensionless
impedances of primary and secondary circuits are plotted in Fig. \ref{cap:zimped}(a)
for the secondary current of the same amplitude but $\pi/2$ phase
lag relative to the primary one. As seen, the results of the present
approach are in good agreement with those of the boundary impedance
condition (BIC) approximation which becomes applicable at sufficiently
high frequencies $(\bar{\omega}>10^{2}).$ The results shown in Fig.
\ref{cap:zimped}(a) imply that for $\bar{\omega}\gtrsim30$ the impedance
of the secondary ring can be compensated by an additional impedance
$Z_{2}=R+\frac{1}{i\bar{\omega}C}$ containing active and capacitative
components denoted by $R$ and $C$, respectively. The relative amplitudes
of the secondary current are plotted in Fig. \ref{cap:zimped}(b)
versus the dimensionless frequency for various impedances added in
the circuit of the secondary coil. The solid curve corresponds to
the short-circuited secondary ring. As seen, the capacitance determines
the resonance frequency at which the amplitude of the secondary current
attains a maximum while its phase becomes delayed by $\pi/2$ with
respect to that of the primary current. The resistance is added to
balance the amplitude of the secondary current with that of the primary
one at the resonance frequency.

\section{Summary and conclusions}

We have presented a boundary-integral equation method for the
calculation of poloidal axisymmetric magnetic fields applicable in
a wide range of ac frequencies. The method is based on the vector
potential formulation and uses Green's functions for Laplace and
Helmholtz equations for the exterior and interior of conductors.
Particular attention was paid to the calculation of Green's
function for the Helmholtz equation which underlies our approach.
In contrast to the Laplace equation, there is no simple analytic
solution for the axisymmetric Green's function of the Helmholtz
equation. Thus, the corresponding function as well as its gradient
has to be calculated numerically that is done by three different
approaches depending on the parameter range. For low and high
dimensionless frequencies we use power series expansions in terms
of elliptical integrals and asymptotic series in terms of modified
Bessel functions, respectively. For the intermediate frequency
range, Gauss-Chebyshev-Lobatto quadratures are used.

Our way of calculation of the Green's function differs
considerably from previous approaches. Note that, on the one hand,
our derivation of the axisymmetric Green's function is more
straightforward and leads to considerably simpler analytic
expressions compared to the Fourier series representation in terms
of Bessel/Hankel and Legendre functions obtained by variable
separation in spherical coordinates \cite{Kost} or to the
Fourier/Bessel integrals obtained by the corresponding integral
transforms in the cylindrical coordinates
\cite{Dodd-Deeds-1968,Yi-Lee-1984}. Fourier series and the
corresponding integrals are computationally more expensive because
they contain products of special functions of a varying argument
whereas the power and asymptotic series in our case contain only a
single special function of a fixed argument and varying order
which can efficiently be calculated using recursion. Moreover, the
selective calculation of the Green's function by either numerical
quadratures, power or asymptotic series depending on its argument
provides the best convergence in each parameter range and, thus,
it is obviously more efficient numerically than a general Fourier
series or corresponding integrals. On the other hand, the way in
which we calculate the Green's function differs significantly from
the approach of Huang \textit{et al.} \cite{Huang} who evaluate
integral (\ref{eq:g-lambda}) numerically. Although before
integration they subtract the Green's function for the Laplace
equation in order to remove the singularity from the integrand,
the discontinuities still remain in derivatives and deteriorate
the accuracy of the numerical integration when the observation
point approaches the contour of integration. Additional
difficulties with numerical integration arise at high
dimensionless frequencies when the exponential function in the
integrand (\ref{eq:g-lambda}) decays in a fast and oscillatory
way.

We avoid these problems by calculating the Green's function analytically
in the form of power and asymptotic series for low and high frequencies,
respectively.

The method was verified by comparison with the analytic solution
for a sphere in a uniform ac magnetic field. In addition, the
performance of the method was demonstrated for a composite model
inductor supplied with a current of fixed amplitude and containing
a secondary coil with an external circuit. In this case, the
results were checked by comparison with an approximate solution
obtained by the boundary impedance condition which becomes
applicable at sufficiently high ac frequencies. The accuracy of
the numerical solution deteriorates at very low frequencies where
an increase of the number of boundary elements is necessary to
obtain a smooth distribution of the magnetic field component
shifted in phase by $\pi/2$ with respect to the applied potential.

The proposed method is well suited for the numerical calculation of
axisymmetric poloidal magnetic field inductors of complicated geometrical
configurations at intermediate ac frequencies because it requires
only the surface but no spatial discretization.

\end{document}